# PHASE–TRANSITION THEORY OF INSTABILITIES. III. THE THIRD–HARMONIC BIFURCATION ON THE JACOBI SEQUENCE AND THE FISSION PROBLEM

Dimitris M. Christodoulou[1], Demosthenes Kazanas[2], Isaac Shlosman[3,4], and Joel E. Tohline[5]

## ABSTRACT

In Papers I and II, we have used a free–energy minimization approach that stems from the Landau–Ginzburg theory of phase transitions to describe in simple and clear physical terms the secular and dynamical instabilities as well as the bifurcations along equilibrium sequences of rotating, self–gravitating fluid systems. Based on the physical picture that emerged from this method, we investigate here the secular and dynamical third–harmonic instabilities that are presumed to appear first and at the same point on the Jacobi sequence of incompressible zero–vorticity ellipsoids.

Poincaré (1885) found a bifurcation point on the Jacobi sequence where a third–harmonic mode of oscillation becomes neutral. A sequence of pear–shaped equilibria branches off at this point but this result does not necessarily imply secular instability. The total energies of the pear–shaped objects must also be lower than those of the corresponding Jacobi ellipsoids with the same angular momentum. This condition is not met if the pear–shaped objects are assumed to rotate uniformly. Near the bifurcation point, such uniformly rotating pear–shaped objects stand at higher energies relative to the Jacobi sequence (e.g. Jeans 1929). This result implies secular instability in pear–shaped objects and a return to the ellipsoidal form. Therefore, assuming that uniform rotation is maintained by viscosity, the Jacobi ellipsoids continue to remain secularly stable (and thus dynamically stable as well) past the third–harmonic bifurcation point.

Cartan (1924) found that dynamical third–harmonic instability also sets in at the Jacobi–pear bifurcation. This result is irrelevant in the case of uniform rotation because the perturbations used in Cartan's analysis carry vorticity and, by Kelvin's theorem of irrotational motion, cannot cause instability. Such vortical perturbations cause differential rotation that cannot be damped since viscosity has been assumed absent from Cartan's equations. Thus, Cartan's instability leads to differentially rotating objects and not to uniformly rotating pear–shaped equilibria. Physically, this instability is not realized in viscous Jacobi ellipsoids because the vortical modes disappear in the presence of any amount of viscosity (cf. Narayan, Goldreich, & Goodman 1987).

---

[1]Virginia Institute for Theoretical Astronomy, Department of Astronomy, University of Virginia, P.O. Box 3818, Charlottesville, VA 22903

[2]NASA Goddard Space Flight Center, Code 665, Greenbelt, MD 20771

[3]Gauss Foundation Fellow

[4]Department of Physics & Astronomy, University of Kentucky, Lexington, KY 40506

[5]Department of Physics & Astronomy, Louisiana State University, Baton Rouge, LA 70803

The fourth–harmonic bifurcation on the Jacobi sequence leads to the dumbbell equilibria that also have initially higher total energies (Paper II). From these considerations, we deduce that a Jacobi ellipsoid can evolve away from the sequence only via a discontinuous $\lambda$–transition (Paper II), provided there exists a branch of lower energy *and broken topology* in any of the known bifurcating sequences. The breaking of topology circumvents Kelvin's theorem and allows a zero–vorticity Jacobi ellipsoid to abandon the sequence.

A pear–shaped sequence has been obtained numerically by Eriguchi, Hachisu, & Sugimoto (1982). Using their results, we demonstrate that the entire sequence exists at higher energies and at higher rotation frequencies relative to the Jacobi sequence. These results were expected since they were predicted by the classical analytical calculations of Jeans (1929). Furthermore, the computed pear–shaped sequence terminates prematurely above the Jacobi sequence due to equatorial mass shedding and thus has no lower energy branch of broken topology. Therefore, there exists no $\lambda$–transition associated with the pear–shaped sequence.

In this case, the first $\lambda$–transition on the Jacobi sequence is of type 2 and appears past the higher turning point of the dumbbell–binary sequence. This transition has been described in Paper II. The Jacobi ellipsoid undergoes fission on a secular time scale and a short–period binary is produced. The classical fission hypothesis of binary–star formation of Poincaré and Darwin is thus feasible. In all stages, evolution proceeds quasistatically and thus the resulting fission is *retarded* just as was anticipated by Tassoul (1978). Modern approaches to the fission problem (Lebovitz 1972; Ostriker & Bodenheimer 1973), involving perfect fluid–masses, are also discussed briefly in the context of phase transitions.

The above conclusions can be strengthened by a more accurate computation of the pear–shaped sequence and by hydrodynamical simulations of viscous Jacobi ellipsoids prior to and past the $\lambda$–point of the dumbbell–binary sequence.

*Subject headings:* galaxies: evolution – galaxies: structure – hydrodynamics – instabilities – stars: formation

# 1  INTRODUCTION

## 1.1  *General Remarks*

Investigations of equilibrium properties and stability characteristics of rotating, self–gravitating fluid masses began in 1687 when Newton introduced his dynamical theory of gravitational attraction and have not ceased throughout the next 308 years as this paper and its companions Paper I and Paper II might indicate. Newton's original motivation was to relate the ellipticity to the rotation of the Earth. During the following centuries, the subject was elevated to new dimensions by various imaginative researchers who envisioned further exciting applications such as to the structure and evolution of single stars, the formation of binary stars and planetary satellites, and the dynamical evolution of the solar system. But it was not until this century when a rigorous treatment by Chandrasekhar (1969) gave us a concrete and unified account of the entire subject through the use of the tensor virial method that is detailed in his monograph "Ellipsoidal Figures of Equilibrium" (hereafter referred to as EFE). In the historical introduction of EFE, we read about how important mathematical tools such as bifurcation theory were developed from the attempts of Poincaré and others to formulate the "fission theory" of rotating, self–gravitating fluid masses and about how these attempts reached an impasse (a colorful term due to Lebovitz 1972) at the bifurcation point of the pear–shaped sequence on the Jacobi sequence. In 1924 Cartan showed that, unlike the bifurcation point of the Jacobi ellipsoids on the Maclaurin sequence, this point is also a point of dynamical instability. "And at this point the subject quietly went into a coma" (EFE).

In the aftermath of our work, we still feel surprised today in front of such a reaction. Instead of questioning immediately this strange coincidence between secular and dynamical instability points (the only one ever found) and instead of trying to understand physically why it occurs, virtually all researchers chose to accept it as a natural result. As far as we are aware, no published work in nearly a century has even discussed *physically* this coincidence (cf. Lyttleton 1953; EFE). Lyttleton (1953), for example, has conveyed the understanding of his time by giving a *mathematical* discussion of dynamical instability that is based on the overall sign of the product of the squares of all oscillation frequencies: dynamical instability is manifested when this product changes sign indicating that the square of one frequency has become negative. This discussion sheds no light to the physics of the modes of dynamical instability. It merely repeats in technical language that the appearance of an imaginary eigenvalue implies instability. Therefore, it is not surprising to find that some of the conclusions obtained for the Jacobi–pear bifurcation turn out to be incorrect simply because the problem has not been defined clearly. Specifically, it has not been clear to the previous investigators that viscosity exists in Jacobi ellipsoids (after all these ellipsoids are products of viscous evolution) and it will damp out any applied vortical perturbations. We provide the relevant details in §2 below but we point out here that our conclusions are similar to those obtained in linear stability analyses of differentially rotating flows which have been performed in order to study the behavior of small–amplitude nonaxisymmetric oscillatory modes about an equilibrium state (Goldreich, Goodman, & Narayan 1986, §5; Narayan, Goldreich, & Goodman 1987, §2).

Based on the results presented for fluids in Paper I and in Paper II, we believe that points of secular and dynamical instability cannot coincide in the case where uniform rotation is assumed. The differences between the two instabilities in fluids are now clear: either a particular conservation law (e.g. angular momentum or circulation) is violated by some physical process leading to secular instability in a "viscous" fluid or all conservation laws are valid (e.g. in a "perfect" nonrelativistic fluid) rendering secular instability irrelevant and leading to dynamical instability at a different point. In both cases, evolution corresponds to a second–order phase transition or a $\lambda$–transition and proceeds toward a sequence that does (Paper I and §4.2 of Paper II) or does not (§2 and §3 of Paper II) bifurcate at the point of instability. It is obvious from this description that points of secular and dynamical instability can coincide only if the two instabilities lead to two distinct sequences that may or may not both bifurcate at the same point. In such a case, one needs to determine which instability is relevant according to the behavior of the conservation laws (see also §1.2 below). [This is also the answer for the "simultaneous" presence of secular and dynamical instability of a test–particle orbit in a central force field of the form $F(r) \propto -1/r^n$, where $r$ denotes distance and $n \geq 3$. Lyttleton (1953, §II) presented this problem as an example of the appearance of both instabilities.]

As Paper I indicates, our work was initially focused on points of instability along the very rich Maclaurin sequence. The main objectives were to understand these instabilities physically and to investigate how they might relate to the controversial Ostriker–Peebles (1973) criterion for global stability of axisymmetric stellar systems. The new mathematical tool at our disposal was the Landau–Ginzburg theory of phase transitions (see Landau & Lifshitz 1986). This approach toward understanding instabilities and bifurcations along equilibrium sequences was originally introduced by Bertin & Radicati (1976) for the Maclaurin–Jacobi bifurcation point and was later extended to more points by group–theoretical methods (Constantinescu, Michel, & Radicati 1979) and with the help of numerical computations (Hachisu & Eriguchi 1983, 1984a). It is also clear from the general descriptions of secular and dynamical stability of Lamb (1932), Lyttleton (1953), and Lebovitz (1977) that the power and simplicity of this approach had been appreciated, but not utilized or properly exploited, in the past.

After using the Maclaurin sequence to gain a clear understanding of the problem and to test the power of the method of solution, we began investigating the Jacobi sequence. We found that the bifurcation of the dumbbell–binary sequence could be understood by analogy to the bifurcation of the axisymmetric one–ring sequence from the Maclaurin sequence (Paper II). Furthermore, as we mentioned above, we realized that the classical results from linear analyses concerning the existence of third–harmonic secular and dynamical instability at the Jacobi–pear bifurcation have been misinterpreted. Seventy years after the work of Cartan (1924), we find that there is no impasse at the Jacobi–pear bifurcation in the classical case where viscosity enforces uniform rotation. In fact, all previously obtained results unanimously support the conclusion that the pear–shaped sequence of objects in uniform rotation is *not at all* accessible from the Jacobi sequence. In this paper, we present the relatively simple physical and mathematical arguments that support these statements.

### 1.2 Basic Assumptions and Dimensionless Quantities

Phase–transition theory allows us to understand all points of bifurcation and instability along well–known equilibrium sequences, such as the Maclaurin and the Jacobi sequences, as points where phase transitions of various orders may appear and may either be *allowed* or may be *forbidden* depending on the (non)–conservation of four "integrals of motion." These integrals of motion are the mass, the energy, the angular momentum, and the circulation.

The most important contribution of this series of papers is the realization that the above integrals of motion determine alone whether/where points of secular and dynamical instability exist. In our terminology, integrals of motion determine whether a phase transition appears at a point and whether a known phase transition is allowed or forbidden. This means that an accurate description of an evolutionary path requires knowledge of whether any integral of motion varies during evolution and understanding of how such an integral may vary in time.

Consider, for example, the pear–shaped ("daughter") sequence that bifurcates from the Jacobi ("mother") sequence but is secularly unstable at its beginning (Lyttleton 1953 and references therein). This simply means that the pear–shaped sequence stands, at least initially, at higher energies relative to the Jacobi sequence (see §2 and §3 below) and implies that no transition is allowed from the mother to the daughter sequence. A simple and clear conclusion can thus be inferred immediately in the case of uniform rotation: the mother sequence continues to be secularly stable to third–harmonic deformations beyond the bifurcation point. This is exactly the opposite conclusion relative to the common belief that the mother and the daughter sequence are both secularly unstable beyond the bifurcation point. This belief is paradoxical; if the Jacobi ellipsoid is secularly unstable to third–harmonic deformations but the pear–shaped sequence is not realized in nature because it is also energetically unstable (Lyttleton 1953, §I), then where is the "unstable" ellipsoid supposed to evolve to? The resolution of the paradox is that, in the complete absence of viscosity, the assumption of uniform rotation is no longer justified and an inviscid Jacobi ellipsoid can evolve toward a differentially rotating state. Cartan's (1924) dynamical instability may indeed indicate such an evolution. The modes studied by Cartan and shown in EFE are, however, vortical (see the analysis of such discontinuous modes in a perfect fluid performed by Case 1960) and the slightest amount of viscosity eliminates them (Narayan, Goldreich, & Goodman 1987; hereafter NGG). For this reason, these modes play no role in determining the evolution of viscous Jacobi ellipsoids which continue to be secularly stable and, thus, necessarily dynamically stable to third–harmonic deformations past the Jacobi–pear bifurcation point. (This example illustrates the state of confusion that the subject is in because no attention has been paid to the behavior of the above integrals of motion and to the role of viscosity. We provide more details in §2.1 below.)

We have encountered an analogous situation in Paper II where we saw that additional sequences of differentially rotating objects may begin at the bifurcation points of the one–ring and dumbbell–binary sequences of uniformly rotating objects, complicating considerably the details of evolution away from the corresponding mother sequences. We hope to investigate the role of differential rotation in Riemann ellipsoids in the future. But for now, it is crucial to retain the assumption of uniform rotation in order to understand the classical results (e.g. Poincaré 1885; Cartan 1924; Jeans 1929; Lyttleton 1953; EFE) concerning third–harmonic perturbations on the Jacobi sequence.

In what follows, we thus consider for simplicity only the evolution of incompressible, uniformly rotating fluid–masses that collapse under the action of self–gravity. We also adopt again the assumptions made in Paper II: (1) Mass is always strictly conserved not only between the initial state and the final equilibrium state but also in the intermediate states that an evolving fluid–mass passes through. (2) Energy always decreases as a fluid–mass evolves toward a new equilibrium state. (3) Angular momentum, expressed in physical units, is always strictly conserved but, in normalized units, it increases or stays constant in time. This general increase in normalized angular momentum $j$ reflects a general increase of the density $\rho$ during the evolution of a contracting fluid–mass [$j \propto \rho^{1/6}$ in equation (1.2) below]. (4) Circulation is strictly conserved only in the complete absence of "viscosity" (i.e. in a "perfect" fluid) but its absolute value decreases during the evolution of a "viscous" fluid–mass. Related to this assumption is the fact that, although the Jacobi ellipsoids have zero vorticity, they are still subject to viscous processes when they are perturbed. This implicit viscosity has no effect on the uniformly rotating equilibrium state but limits considerably the spectrum of the physically important perturbations (see §2 below).

In §3, we analyze results from computations of the pear–shaped sequence, obtained by Eriguchi, Hachisu, & Sugimoto (1982), using the dimensionless quantities $\omega$ (rotation frequency), $j$ (angular momentum), and $E$ (total energy) that were introduced in Paper II. These quantities are defined by

$$\omega^2 \equiv 10^2 \Big(\frac{\Omega^2}{4\pi G \rho}\Big), \tag{1.1}$$

$$j^2 \equiv 10^2 \Big(\frac{L^2}{4\pi G \rho^{-1/3} M^{10/3}}\Big), \tag{1.2}$$

and

$$E \equiv 10^4 \Big[\frac{T+W}{(4\pi G)^2 M^5 L^{-2}}\Big], \tag{1.3}$$

where $\Omega$ is the rotation frequency, $G$ is the gravitational constant, $\rho$ is the mass density, $L$ is the angular momentum, and $M$ is the mass, all in physical units. The term $T+W$ denotes the total energy in physical units as the sum of the total kinetic energy due to rotation $T$ and the total gravitational potential energy $W$. As we have seen in Paper I, the total energy $T+W$ and its dimensionless counterpart $E$ also represent the free-energy function when they are not constrained by equilibrium conditions (for more details see Tohline & Christodoulou 1988). Finally, when $\omega$, $j$, and $E$ refer to equilibrium configurations specifically we write them as $\omega_o$, $j_o$, and $E_o$, respectively.

### 1.3 Outline

The remainder of the paper is organized as follows. In §2, we discuss the Jacobi–pear bifurcation and the related results from linear analysis obtained by Cartan (1924), Lyttleton (1953), Chandrasekhar (EFE), and others. Cartan's dynamical instability does not appear in the presence of a slight amount of viscosity that eliminates vortical perturbations (cf. NGG). So, in the classical case of uniform rotation maintained by viscosity, Cartan's result is irrelevant and the "third–harmonic instabilities" are not realized because the pear–shaped sequence stands at higher energies relative to the Jacobi sequence (Lyttleton 1953 and references therein). This result implies that the Jacobi sequence remains stable beyond the third–harmonic bifurcation point.

In §3, we turn to the numerical results of Eriguchi, Hachisu, & Sugimoto (1982) who have computed the physical properties of the bifurcating pear–shaped sequence. Using their tabulated results, we demonstrate that the entire pear–shaped sequence stands at higher free energy (and at higher rotation frequency as well) relative to the Jacobi sequence. We conclude then that the Jacobi ellipsoids do not suffer a secular or a dynamical third–harmonic instability as Poincaré (1885), Cartan (1924), Lyttleton (1953), Chandrasekhar (EFE), and many others believed. Therefore, the only known process that can drive the evolution of a Jacobi ellipsoid away from its sequence is a topology–breaking $\lambda$–transition (see Paper II). However, the pear–shaped sequence terminates prematurely above the Jacobi sequence due to equatorial mass shedding and has no lower energy branch like the dumbbell–binary sequence (Paper II). This result implies that, unlike in the case of the dumbbell–binary sequence, there exists no $\lambda$–transition associated with the pear–shaped sequence. These results point to a re–examination of the classical fission hypothesis of Poincaré and Darwin (see EFE and Tassoul 1978).

In §4, we revisit the fission problem. We modify the last stages of the classical fission hypothesis of Poincaré and Darwin in light of our results. Like the classical fission hypothesis, the new scenario of binary fission postulates secular evolution driven by viscosity but, in addition, it does not fail at the Jacobi–pear bifurcation and incorporates the $\lambda$–transition of the dumbbell–binary sequence that finally drives a Jacobi ellipsoid away from its sequence. This scenario is free of unfounded speculations and, we believe, free of errors. We argue that this *modified classical fission hypothesis* leads necessarily to *retarded fission* — as was anticipated by Tassoul (1978) — because the entire evolution takes place on a secular time scale. Thus, this modified scenario explains the formation of short–period, spectroscopic binaries. The modern approaches of Lebovitz (1972, 1974, 1987) and of Ostriker (e.g. Ostriker & Bodenheimer 1973) to the fission problem are also discussed briefly in this section (see also the reviews by Lebovitz 1977, Tassoul 1978, and Durisen & Tohline 1985).

In §5, we summarize our results and we discuss their implications.

## 2  The Perturbed Jacobi Ellipsoids

### 2.1  Previous Work and Secular Stability

The fluid–dynamics researchers of the previous century understood well the physical importance of irrotational motion. Lamb (1932, §33) wrote the following on this subject:

"The physical importance of the subject rests on the fact that if the motion of any portion of a fluid mass be irrotational at any one instant it will under certain very general conditions continue to be irrotational....It follows that if the motion of any portion of a fluid mass be initially irrotational it will always retain this property; for otherwise the circulation in every infinitely small circuit would not continue to be zero, as it is initially..."

It is thus surprising that those who performed linear stability analyses on the Jacobi sequence of ellipsoids did not pay attention to the property that makes these systems different than all the other Riemann ellipsoids: the Jacobi ellipsoids have zero vorticity when they are viewed in a rotating coordinate frame in which their firgures appear stationary. It is also surprising that a celebrated theorem of fluid dynamics due to Kelvin which is well–respected in modern analyses of hydrodynamical flows (e.g. Goldreich, Goodman, & Narayan 1986, §5, hereafter GGN; see also Landau & Lifshitz 1987) has also been ignored during the attempts to investigate the dynamical stability of Jacobi ellipsoids to third–harmonic perturbations. This theorem states that the irrotational motion of a liquid occupying a simply–connected region has less kinetic energy than any other motion consistent with the same normal motion of the boundary (e.g. Lamb 1932, §45). Kelvin's theorem implies that if the initial equilibrium state has no vorticity, then only nonvortical perturbations are important for stability since vortical perturbations raise the kinetic energy and, therefore, cannot cause instability. Physically, such vortical perturbations are eliminated by the slightest amount of viscosity. [See the analysis of Case (1960) but also the description of NGG who, faced with a problem similar to that of Cartan's (1924), eliminated such modes from consideration and successfully concentrated on the remaining physical modes.]

The lack of attention to the fundamental theorems of fluid dynamics stated above has led to confusion in the interpretations of the results concerning third–harmonic perturbations on the Jacobi sequence. All these previous results are summarized by Lyttleton (1953) and, to a lesser extent, in EFE. In particular, a crucial proof that the pear–shaped objects near the Jacobi–pear bifurcation are secularly unstable was ignored in favor of Cartan's (1924) "proof of dynamical instability." The secular instability of the pear–shaped sequence was discovered by Liapounoff and was later confirmed by Jeans (both cited in Lyttleton 1953 and in Lebovitz 1977; see also §3 below). This result implies that the Jacobi ellipsoids are secularly stable (and, thus, also dynamically stable due to lack of available lower energy states) to third–harmonic perturbations and, therefore, points to the fact that Cartan's (1924) "dynamical instability" does not exist (see §2.2 below). In fact, Lyttleton (1953, §VIII) has described clearly and without errors the return of a pear–shaped object back to the Jacobi ellipsoidal form after dissipation of the excess energy. It is unfortunate that Lyttleton, who obviously understood well the secular instability of the pear–shaped objects, gave too much weight to the calculation of Cartan (1924) and was finally led to the erroneous conclusion that the Jacobi ellipsoids must also become secularly unstable beyond the Jacobi-pear bifurcation. As we mentioned above, this cannot be the case because the pear–shaped sequence is itself unstable and sits at higher energies relative to the Jacobi sequence. We conclude that the Jacobi ellipsoids do not become secularly unstable since there is no lower energy state available to them. Furthermore, secular stability also implies automatically dynamical stability.

The above classical results have been confirmed implicitly by Eriguchi, Hachisu, & Sugimoto (1982). In §3 below, we use their tabulated numerical results to demonstrate explicitly that the entire pear–shaped sequence is unstable because it exists at higher energies and at higher rotation frequencies relative to the Jacobi sequence. This behavior of the sequence in the $(j_o^2, \omega_o^2)$ plane may seem surprising at first but, in fact, it was anticipated because of the results of Jeans (1929) who has proved that the pear–shaped figures near the Jacobi–pear bifurcation exist at higher rotation frequencies relative to the unperturbed Jacobi ellipsoids. These results are described also by Lyttleton (1953) and in §3 below.

## 2.2 The Jacobi–Pear Bifurcation and Dynamical Stability

We have seen in §2.1 that the Jacobi sequence is secularly stable and, thus, dynamically stable as well, to third–harmonic perturbations. The question then is: What is the nature of Cartan's (1924) "dynamical instability" at the Jacobi–pear bifurcation?

In the absence of viscosity, Cartan's instability appears at the third–harmonic bifurcation because the neutral mode (EFE) that deforms the critical Jacobi ellipsoid to a pear–shaped object happens to conserve circulation automatically. This can be shown easily in Cartesian coordinates $x, y, z$ (referring to axes $X, Y, Z$) by a direct integration that obtains the surface area $A_P$ of the perturbed object under a closed curve at any plane $P$ perpendicular to the rotation axis $Z$. $A_P$ turns out to be equal to the corresponding surface area of the undeformed Jacobi ellipsoid because the proper displacement is independent of the $x$–coordinate and is applied only in the $X$–direction [see EFE, §40, equation (58)]. Because the rotation frequency $\Omega$ is also unchanged to first order by the deformation (Lyttleton 1953; EFE), the circulation $C_P \equiv -2\Omega A_P$ is conserved by the applied proper displacement under any closed curve at any plane $P$.

For true dynamical instability, circulation should also be conserved beyond the bifurcation point by the applied perturbations which must then be nonvortical. However, the adopted perturbations carry vorticity violating circulation conservation and Kelvin's theorem of irrotational motion given in §2.1 above. This is also relatively easy to show and we proceed to do so below.

The stability of incompressible inviscid Jacobi ellipsoids to infinitesimal third–harmonic perturbations was examined by Cartan (1924) using the the hydrodynamical equations for small oscillations. The method of solution is described in detail by Lyttleton (1953). (Cartan's result was confirmed in EFE by an application of the tensor virial method but it will be clear below that both approaches suffer from the same inconsistency.) In a Cartesian coordinate frame $XYZ$ with corresponding coordinates $x, y, z$ and rotating with constant frequency $\Omega$ (the rotation frequency of the Jacobi ellipsoid in the inertial frame) about the $Z$ axis, these equations are the three components of the equation of motion

$$\frac{\partial u}{\partial t} - 2\Omega v = -\frac{\partial \Psi}{\partial x}, \tag{2.1}$$

$$\frac{\partial v}{\partial t} + 2\Omega u = -\frac{\partial \Psi}{\partial y}, \tag{2.2}$$

$$\frac{\partial w}{\partial t} = -\frac{\partial \Psi}{\partial z}, \tag{2.3}$$

and the continuity equation

$$\frac{\partial u}{\partial x} + \frac{\partial v}{\partial y} + \frac{\partial w}{\partial z} = 0, \tag{2.4}$$

where $t$ denotes time, $\mathbf{v}=(u, v, w)$ is the velocity vector in the rotating frame, and $\Psi$ is the effective potential given by

$$\Psi = Q + \Phi - \frac{1}{2}\Omega^2(x^2 + y^2). \tag{2.5}$$

The symbols $Q$ and $\Phi$ denote the enthaply and the gravitational potential, respectively.

Assuming normal modes of the general form

$$\mathcal{F}(x, y, z, t) = \mathcal{F}(x, y, z)e^{i\sigma t}, \tag{2.6}$$

where $\mathcal{F}$ represents any of the components $u, v, w$, the eigenfrequencies $\sigma$ are sought for which equations (2.1)–(2.4) have a solution that satisfies the boundary condition that $Q = 0$ on the perturbed surface of the object. A mode with $\sigma = 0$ is neutral since the amplitudes $\mathcal{F}$ do not depend on time; a mode with $\sigma^2 < 0$ is dynamically unstable since all amplitudes grow exponentially in time; and a mode with $\sigma^2 > 0$ is stable since all amplitudes oscillate periodically in time.

The above description of normal–mode linear stability analysis contains all the physics of the problem. The rest of the calculation of Cartan (1924) is involved and technical only because it is concerned with the determination of the gravitational potential $\Phi$ on the deformed boundary surface. (This problem is solved elegantly in EFE through the use of the tensor virial equations.)

Equations (2.1)–(2.4) are sufficient for the purposes of our analysis. Substitution of the normal modes (2.6) into these equations leads to Poincaré's equation

$$\frac{\partial^2 \Psi}{\partial x^2} + \frac{\partial^2 \Psi}{\partial y^2} + \left(1 - \frac{4\Omega^2}{\sigma^2}\right)\frac{\partial^2 \Psi}{\partial z^2} = 0, \tag{2.7}$$

which is the fundamental equation in Cartan's (1924) analysis. This equation describes all the normal modes of oscillation induced by a particular deformation of the boundary surface. We shall find it useful to describe some of the intemediate steps in obtaining Poincaré's equation. Substituting the form (2.6) into (2.1)–(2.3) and differentiating each equation once with respect to its spatial variable on the right–hand side, and then using equation (2.4), one obtains

$$\nabla^2 \Psi = 2\Omega(\nabla \times \mathbf{v}) \cdot \mathbf{k} = 2\Omega\zeta, \tag{2.8}$$

where $\mathbf{k}$ is the unit vector along the $Z$ axis of the coordinate frame and the term $\zeta \equiv (\nabla \times \mathbf{v}) \cdot \mathbf{k}$ is the $Z$–component of the vorticity vector in the rotating frame. Equation (2.8) is a Poisson equation in which $\zeta$ serves as the source of the effective potential $\Psi$.

Differentiating next equations (2.1) and (2.2) with respect to $y$ and $x$, respectively, and using equations (2.6) and (2.4), one finds that

$$i\sigma\zeta = 2\Omega\frac{\partial w}{\partial z}, \tag{2.9}$$

or, dividing by $i\sigma$ (an assumption that is fully justified in NGG and, as they explain, also restricts the analysis to perturbations with no $z$–dependence — see below) and using equations

(2.6) and (2.3), that
$$\zeta = \frac{2\Omega}{\sigma^2} \frac{\partial^2 \Psi}{\partial z^2}. \tag{2.10}$$

Poincaré's equation (2.7) follows by eliminating $\zeta$ between equations (2.8) and (2.10) but this elimination of $\zeta$ is the source of inconsistency because the perturbations are then allowed to generate vorticity. This cannot be seen explicitly from Poincaré's equation (2.7) but, nevertheless, violates Kelvin's theorem of irrotational motion (§2.1) as well as circulation conservation in Jacobi ellipsoids. Such vortical perturbations were studied by Case (1960) in two–dimensional Couette flow of a perfect fluid and were dismissed by NGG in their analysis of the shearing sheet. Without going into details about the nature of these perurbations (which can be found in the above references), we note once again that such perurbations are not important for evolving Jacobi ellipsoids because they are eliminated by an infinitesimal amount of viscosity.

It is clear from the above description that third–harmonic perturbations cannot cause any instability in Jacobi ellipsoids because they cannot be nonvortical maintaining $\zeta = 0$. In fact, this is true on the Jacobi sequence for all higher–order surface deformations as well since the analysis that leads to equation (2.10) applies to all of these cases. In this sense, dynamical instability cannot occur anywhere on the Jacobi sequence because, as long as topology does not break, secular instability does not appear; obeying Kelvin's theorem, all the bifurcations of the Jacobi sequence must lead initially to sequences of higher energy. The critical Jacobi ellipsoids at the various bifurcation points are secularly stable simply because they have no more vorticity to lose. This simple observation has escaped the attention of researchers for more than a century.

Another way of showing that there can be no secular instability at the bifurcation points of the Jacobi sequence was hinted in Paper II. Let us ignore, for the sake of the argument, lower energy branches that are accessible in any case only via a discontinuous $\lambda$–transition (see Paper II). Then, assuming that there is a daughter sequence of deformed objects that bifurcates from the Jacobi sequence toward lower energies is contradictory because it implies that there exist intermediate *nonequilibrium* states of progressively lower energy *and lower vorticity* analogous to Riemann S–type ellipsoids — but the vorticity of the uniformly rotating Jacobi ellipsoids in the rotating coordinate frame cannot become lower than zero.

In order to examine physically acceptable, nonvortical perturbations in *inviscid Jacobi ellipsoids* one must set $\zeta = 0$ in equations (2.8) and (2.10). In this case, the eigenvalue problem changes drastically from Cartan's (1924) problem. These two equations take the forms
$$\nabla^2 \Psi = 0, \tag{2.11}$$
and
$$\frac{\partial^2 \Psi}{\partial z^2} = 0, \tag{2.12}$$
respectively. Combining equations (2.11) and (2.12) one finds that the problem is now described by a Laplace equation of the form $\nabla_2^2 \Psi = 0$, where the symbol $\nabla_2^2$ denotes the Laplacian operator in two dimensions $(x, y)$. The eigenvalue $\sigma$ appears now only in the boundary conditions (cf. Lamb 1932; GGN). The new eigenvalue problem is essentially two–dimensional; equation (2.12) is just a constraint leading to modes with no $z$–dependence since $\Psi$ must be an even function of $z$ under the assumption of equatorial symmetry. This problem is not relevant in the physically interesting case of evolving *viscous Jacobi ellipsoids* and will not concern us here any more (but see the analysis in §5 of GGN).

## 3    The Pear–Shaped Sequence

A sequence of uniformly rotating pear–shaped objects has been obtained numerically by Eriguchi, Hachisu, & Sugimoto (1982). Although their Jacobi sequence plotted in the $(j_o^2, \omega_o^2)$ plane along with the pear–shaped sequence is incorrect, we believe that the tabulated values for the pear–shaped objects are correct and we proceed to interpret these results in this section. [Our belief is based on the analytical results obtained by Jeans (1929) which we also analyze below. It turns out that the tabulated results agree while the plotted results disagree with the analytical predictions of Jeans.]

The tabulated values for the pear–shaped objects are plotted as open circles in Figures 1 and 2 in the $(j_o^2, \omega_o^2)$ plane and in the $(j_o^2, E_o)$ plane. The Jacobi sequence has been calculated analytically and is also plotted as a dashed line. The pear–shaped sequence should bifurcate at point A where the neutral third–harmonic mode of oscillation appears (EFE). The Jacobi ellipsoid at the bifurcation point A has axes ratios of $b/a = 0.432232$ and $c/a = 0.345069$, where the axes $a, b, c$ lie along the axes $X, Y, Z$, respectively, of the rotating coordinate frame. These values correspond to an equatorial eccentricity of $\eta = 0.901762$ and a meridional eccentricity of $e = 0.938577$, respectively, and, in our system of units, to $j_o^2 = 0.749071$, $\omega_o^2 = 7.100757$, $E_o = -4.440868$, and $T/|W| = 0.162836$.

The first important observation in Figures 1 and 2 is that the pear–shaped sequence does not extend to all values of the angular momentum $j_o$. Specifically, the sequence exists only prior to $j_o^2 = 0.8079$. Therefore, all Jacobi ellipsoids with $j_o^2 > 0.8079$ cannot be unstable to third–harmonic perturbations and to a transition to such pear–shaped objects simply because the pear–shaped sequence does not exist at such high angular momentum values.

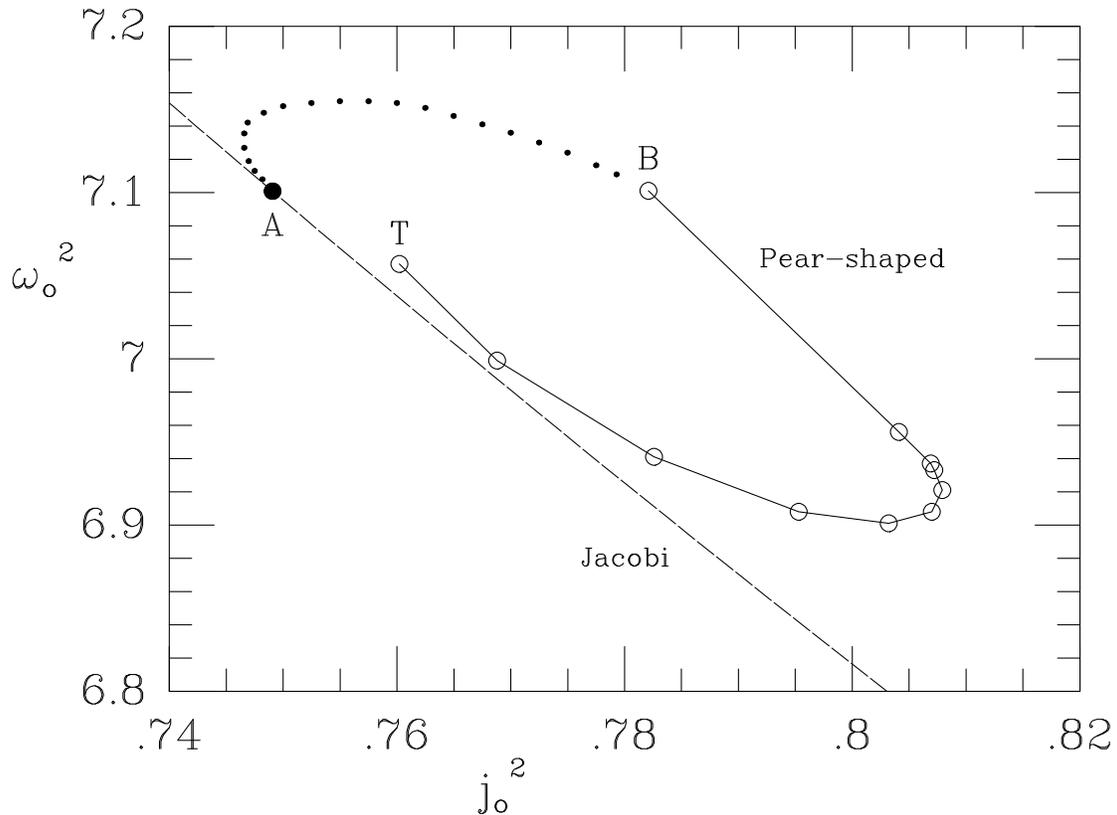

FIGURE 1. *The rotation frequency squared $\omega_o^2$ along the pear–shaped and the Jacobi sequence is plotted as a function of the dimensionless angular momentun squared $j_o^2$. The Jacobi sequence has been calculated analytically and compares very well with the analytical values tabulated by Tassoul (1978) and with the numerical values tabulated by Hachisu & Eriguchi (1982). The exact bifurcation point A (filled circle) has also been calculated analytically using an equatorial axis ratio of $b/a = 0.432232$ (EFE). All points (open circles) of the pear–shaped sequence were obtained from Table I of Eriguchi, Hachisu, & Sugimoto (1982) who gave B as the bifurcation point and T as the termination point due to equatorial mass shedding. The dotted line is based on the analytical calculations of Jeans (1929) and shows schematically the pear–shaped sequence bifurcating toward higher rotation frequencies and lower angular momentum values before turning back toward point B.*

The points denoted as B and T in Figures 1 and 2 were given by Eriguchi, Hachisu, & Sugimoto (1982) as the bifurcation and the termination point, respectively. The pear–shaped sequence terminates because of equatorial mass shedding while all equilibria are still only slightly deformed away from the ellipsoidal form. Whether the sequence terminates at T or at B may seem questionable but does not really affect our interpretations. We believe that T is really the termination point and that the pear–shaped sequence bifurcates not only above the Jacobi sequence but also backwards. Then, it turns around toward higher values of $j_o$ and proceeds to join the solid line in Figure 1 at point B.

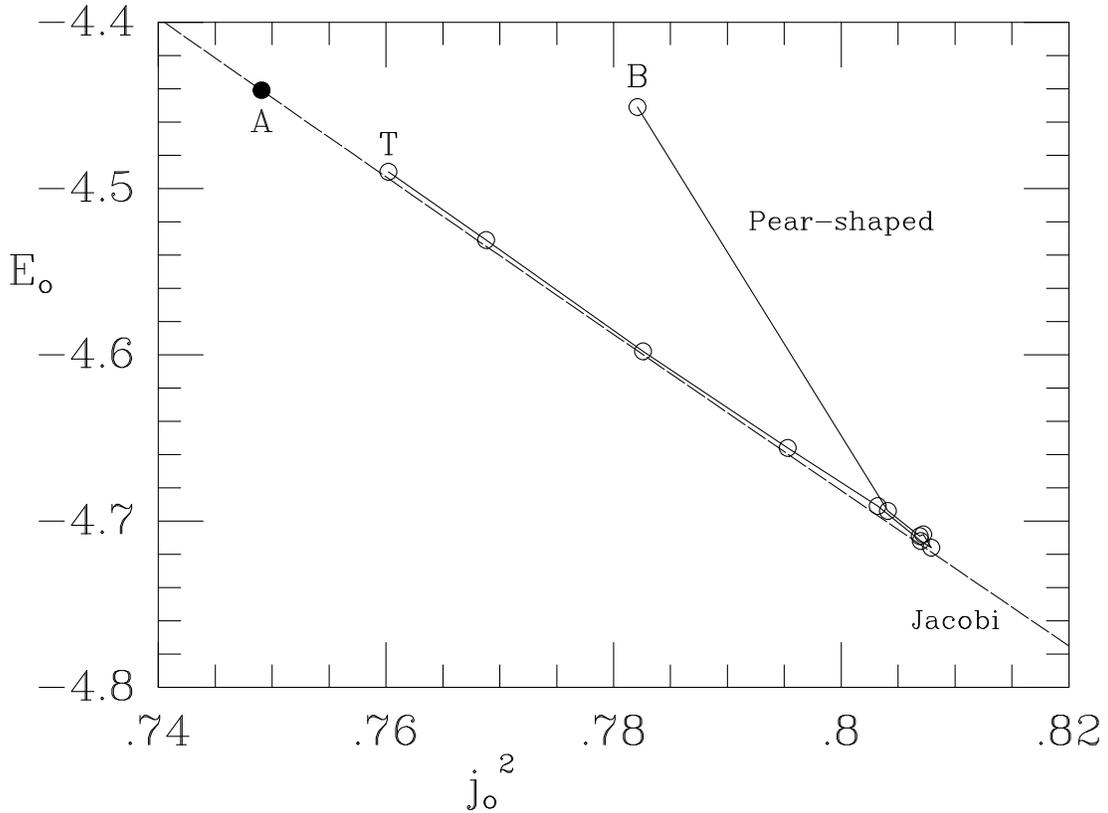

FIGURE 2. The free–energy $E_o$ is plotted versus $j_o^2$ for the sequences of Figure 1.

The path just described is shown schematically in Figure 1 (dotted line) and is based on the analytical results obtained by Jeans (1929) [see also Lyttleton (1953), §VIII, equation (15)] who showed that the pear–shaped objects with higher rotation frequencies relative to the critical Jacobi ellipsoid (that exists at the Jacobi–pear bifurcation) also have lower angular momentum values. Specifically, Jeans showed that the angular momentum $j_p$ and the rotation frequency $\omega_p$ of the pear–shaped equilibria near the bifurcation point $(j_o, \omega_o)$ are given by the equations

$$j_p = j_o(1 - 0.06765\epsilon^2), \tag{3.1}$$

and

$$\omega_p^2 = \omega_o^2(1 + 0.05227\epsilon^2), \tag{3.2}$$

respectively, where $\epsilon << 1$ is a "departure" parameter away from the bifurcation point and the critical ellipsoidal form. It is immediately seen from these equations that $j_p < j_o$ and $\omega_p > \omega_o$. Eliminating $\epsilon$ between equations (3.1) and (3.2), we find that

$$\frac{\Delta\omega^2}{\omega_o^2} = 0.77265\frac{|\Delta j|}{j_o}, \tag{3.3}$$

where $\Delta\omega^2 \equiv \omega_p^2 - \omega_o^2 > 0$ and $|\Delta j| \equiv j_o - j_p > 0$. This equation determines the slope $d\omega_o^2/dj_o^2$ of the schematic dotted line at point A in Figure 1.

The numerical results confirm that the Jacobi ellipsoids suffer neither a secular nor a dynamical instability from third–harmonic perturbations. More importantly, Figures 1 and 2 show explicitly that there is no $\lambda$–transition associated with the pear–shaped sequence since there is no lower energy branch that belongs to this sequence. (Type 1 and type 2 $\lambda$–transitions have been discussed in detail in Paper II.) This result is important because the $\lambda$–transitions of type 2 are discontinuous catastrophes and cannot be detected by linear analyses of the kind performed on the Jacobi sequence and described in EFE. We conclude that an evolving Jacobi ellipsoid does not suffer a third–harmonic instability of any kind, including $\lambda$–transitions; hence, it cannot abandon the sequence during evolution and it cannot be driven toward another equilibrium state by any existing third–harmonic perturbations.

## 4   The Fission Problem Revisited

### 4.1   The Classical Approach to the Fission Problem

The classical fission hypothesis has its origin in the ideas of Poincaré (1885) who was the first to formulate the bifurcation theory of new equilibrium sequences from the Maclaurin and the Jacobi sequences. Poincaré discovered the Maclaurin–Jacobi and the Jacobi–pear bifurcations and also realized that viscosity is necessary for the transformation of an evolving Maclaurin spheroid to a triaxial ellipsoid evolving along the Jacobi sequence beyond the former bifurcation. Furthermore, Poincaré realized the competition between two different time scales during evolution (cf. Lebovitz 1972, 1977), the secular instability time scale determined by viscosity and the contraction time scale determined by the dissipation of energy during nonequilibrium evolution of a fluid–mass. He clearly stated that the viscosity should be large enough and the contraction slow enough for quasistatic evolution to be viable. The first condition secures the appearance of the second–harmonic secular instability at the Maclaurin–Jacobi bifurcation. The second condition allows for evolution to be slow and thus to be described by a series of successive transitions along an equilibrium sequence and from one sequence to another when secular instability sets in.

When Poincaré discovered the Jacobi–pear bifurcation point he assumed that this is also a point of secular instability and imagined that the evolution will switch again and will begin to follow the pear–shaped sequence resulting eventually in a detachment of a fragment from the object at sufficient elongation and in the creation of a satellite. Darwin expanded this idea and effectively introduced the concept of binary–star formation via fission from the pear–shaped sequence (see Lyttleton 1953 and EFE for more details). These evolutionary scenarios became less believable when Cartan (1924) showed that "dynamical instability" sets in at the Jacobi-pear bifurcation and interrupts the quasistatic evolution. On the other hand, the work presented in this paper shows that there is no third–harmonic instability of any kind on the Jacobi sequence and, therefore, points to a re–examination of the classical fission hypothesis and to a modified evolutionary path relative to that imagined by Poincaré and Darwin.

In what follows, we adopt the classical formulation of the fission problem. We assume that an evolving fluid–mass is incompressible, self–gravitating, and isolated from its envi-ronment. Furthermore, we postulate that viscosity is always present in the fluid and is sufficiently strong to enforce uniform rotation at all times.

The "modified classical fission hypothesis" proceeds initially along the path outlined by Poincaré and others. A spheroidal fluid–mass contracts slowly and evolves along the Maclaurin sequence as it radiates some ot its energy away. This type of quasistatic evolution corresponds to a gradual increase of the normalized angular momentum $j_o$ defined in §1.2 above. The evolving fluid–mass becomes secularly unstable to second–harmonic perturbations at the Maclaurin–Jacobi bifurcation and undergoes a symmetry–breaking second–order phase transition. Evolution switches to the Jacobi sequence. The fluid–mass is now ellipsoidal in shape but is still uniformly rotating and has no vortical motions.

No second–harmonic or third–harmonic instability of any kind is encountered on the Jacobi sequence. Attempts by vortical perturbations to induce differential rotation are quickly damped by viscosity which, however, plays no role in defining the equilibrium structure of the ellipsoidal fluid–mass. The fluid–mass then crosses the third–harmonic bifurcation point of the unstable pear–shaped sequence and the fourth–harmonic bifurcation point of the unstable branch of the dumbbell sequence (Paper II) and proceeds to the type 2 $\lambda$–point associated with the higher turning point of the dumbbell sequence. This $\lambda$–point is defined *on the Jacobi sequence* by the $j_o$–value of the latter point (see Paper II for details). The Jacobi ellipsoid at the $\lambda$–point has $j_o^2 = 1.540$, $\omega_o^2 = 4.25864$, $E_o = -7.77253$, and $T/|W| = 0.20236$. Its equatorial and meridional eccentricities are $\eta = 0.97201$ and $e = 0.97743$ corresponding to axes ratios of $b/a = 0.23494$ and $c/a = 0.21126$, respectively.

Further contraction triggers a type 2 $\lambda$–transition (Paper II). The defining characteristic of type 2 $\lambda$–transitions is the ensuing spontaneous breaking of the topology without a breaking of the symmetry of the fluid–mass. This $\lambda$–transition is also a "discontinuous catastrophe" that develops on a time scale determined by viscous dissipation. The ellipsoid breaks up into two equal–mass fragments on a secular time scale and the newborn binary system emerges on the lower energy *stable* branch of the dumbbell–binary sequence.

The entire path that we have described is slow and quasistatic. The resulting fission therefore takes a generally long time to materialize — it is *retarded* (Tassoul 1978). The two individual components of the binary system are close to each other. It is then conceivable that, when the two stars emerge on the main sequence and become easily observable, the system is still a close binary. In this case, the retarded fission mechanism described above may explain the existence of the observed short–period, spectroscopic binaries. [See also Abt (1978, 1983), Abt, Gomez, & Levy (1990), and Bodenheimer (1992) for more details on observed spectroscopic binaries and for related discussions.]

Two questions are important in relation to the above description: (a) How can triple, quadruple, and generally multiple systems be formed since this fission mechanism always leads to binary fission? (b) How can unequal–mass binaries be formed since this fission mechanism always produces equal–mass components? We discuss some possible answers to these questions below. We concentrate on fission mechanisms via $\lambda$–transitions that take place from equilibrium initial conditions (i.e., we do not include other possible mechanisms such as capture and nonequilibrium cataclysmic fragmentation).

Multiple systems do not form by the above mechanism and do not form via quasistatic evolution. Such systems can naturally form via type 3 $\lambda$–transitions that take place on dynamical time scales and originate on the lower, toroidal branch of the one–ring sequence (Eriguchi & Hachisu 1983b; Paper II). These dynamical transitions will be discussed in their proper context in the final paper of this series (hereafter referred to as Paper IV).

Equilibrium sequences of unequal–mass binaries have been calculated numerically by Hachisu & Eriguchi (1984b). These sequences exist in the general area of the $(j_o^2, \omega_o^2)$ plane that is bounded by the absolutely stable parts of the Maclaurin and the Jacobi sequences and by the equal–mass (dumbbell–)binary sequence. Naturally, the mass ratio of the two components increases with decreasing angular momentum $j_o$ and tends to infinity in the limit of the absolutely stable part of the Maclaurin sequence. The unequal–mass binary sequences can be accessed from the absolutely stable parts of the Maclaurin and the Jacobi sequences but only if these objects are disturbed by finite–amplitude perturbations. In such a case, nonlinear breakup may produce isolated unbound fragments or, in some cases, bound unequal–mass binaries because the equilibrium energies along these binary sequences are, for the most part, lower than those of the corresponding Maclaurin spheroids and Jacobi ellipsoids. The conditions for such nonlinear breakup are most favorable at *nonlinear resonances* where the equilibrium energies of the two sequences differ only slightly from each other in the $(j_o^2, E_o)$ plane. Finally, another possible formation mechanism is indirect and begins with the formation of an unequal–mass multiple system via a type 3 $\lambda$–transition from the one–ring sequence. Then, unequal–mass binaries may be left behind after the ejection of the additional low–mass components. Nonlinear resonances and the indirect formation mechanism will be discussed in Paper IV.

### *4.2 Modern Approaches to the Fission Problem*

Besides the classical approach, there have been two more approaches to the fission problem (see Tassoul 1978 for a review of all the previous viewpoints). Both of these "modern approaches" are mainly concerned with compressible fluid masses and adopt the viewpoint that the viscous time scale is much longer than the contraction time scale and, thus, they assume that a contracting fluid–mass may be considered, for all practical purposes, as a perfect fluid in which viscosity is completely absent. Hence, the modern approaches rely on dynamical instabilities to induce fission. We discuss briefly these two modern approaches below.

### *4.2.1 The Approach of Ostriker*

We have already described the approach of Ostriker (e.g. Ostriker & Bodenheimer 1973) to the fission problem in Paper II. This approach does not lead to binary fission because differential rotation takes over when an axisymmetric perfect fluid–mass suffers a second–harmonic dynamical instability. Viscosity has been assumed absent and so there is no mechanism that can damp the differential vortical motions induced by the applied perturbations. The outcome of the evolution away from the perfect–fluid Maclaurin sequence is some kind of core–ring equilibrium (cf. Eriguchi & Hachisu 1983a; see also the simulations performed by Tohline, Durisen, & McCollough 1985, Williams & Tohline 1987, 1988, and the review article of Durisen & Tohline 1985).

As was mentioned in Paper II, Ostriker's idea of cataclysmic fission was just short of the third–harmonic dynamical instability point on the Maclaurin sequence which is a $\lambda$–point of type 3. The $\lambda$–transition at this point does result in binary fission on a dynamical time scale. We shall discuss this nonlinear process in more detail in Paper IV.

### 4.2.2   The Approach of Lebovitz

An interesting scenario for binary fission was proposed and discussed by Lebovitz (1972, 1974, 1977, 1987). In this scenario, viscosity is assumed again to be absent leading to strict conservation of both circulation and angular momentum during evolution. A compressible perfect fluid–mass is perturbed slightly away from the Maclaurin sequence and, thus, avoids the second–harmonic dynamical instability as it evolves through different sequences of compressible Riemann ellipsoids. Assuming that the perturbation amplitudes are relatively small, the fluid–mass is expected to remain close to the Maclaurin line and, subsequently, it will continue to evolve close to the $x = +1$ self–adjoint Riemann sequence. Since all Riemann ellipsoids are stable to second–harmonic perturbations, the fluid–mass is expected to be subject to instabilities induced by perturbations belonging to third or even higher harmonics.

The presence of compressibility and the absence of viscosity introduce important changes in the approach of Lebovitz as compared to the viscous incompressible case of §4.1. From the above discussions, it is easy to see why some viscosity may have to be brought back into this scenario: viscosity can damp out, at least partially, differential motions and vortical perturbations. It is quite possible that a pear–shaped sequence with vorticity bifurcates from each Riemann sequence but such pear–shaped objects will certainly develop differential rotation during evolution. Although it is unknown where such sequences stand relative to the corresponding Riemann sequences, we have no indication that they are unstable especially since these equilibria can have vorticity. Thus, if such pear–shaped objects are stable, some viscosity may be useful in preventing the transitions of Riemann ellipsoids to the pear–shaped sequences and in allowing the evolution to proceed to the region of the parameter space where fourth–harmonic perturbations become important. However, it is also unknown whether compressible Riemann sequences extend that far in the parameter space and this is why compressibility becomes important in the approach of Lebovitz. Finally, it is not known whether conditions for a phase transition will exist in the vicinity of each fourth–harmonic bifurcation. But this seems possible and is not likely to pose a particularly difficult problem compared to the possibility that entire compressible Riemann sequences may terminate prior to the fourth–harmonic bifurcations (see related results in Hachisu & Eriguchi 1982).

Whether fission of compressible fluid–masses is really viable will ultimately be decided on several levels. (1) Equilibrium sequences of compressible triaxial and multi–fluid bodies must be studied in more detail. To this extent, the first attempts to compute properties of such sequences (Hachisu & Eriguchi 1982; Eriguchi & Hachisu 1983b; Hachisu 1986) are very useful. (2) The role of viscosity in damping vortical motions and in preventing transitions to differentially rotating core–ring states is not well–understood and must be further investigated. (3) There currently exists no evidence that compressible fluid–masses can avoid third–harmonic instabilities and transitions to pear–shaped objects which are prone to developing differential motions (but see the attempts of Lebovitz 1974 to investigate this question). (4) Even if third–harmonic instabilities can be avoided, it is not known whether type 2 $\lambda$–transitions or other instabilities exist and are associated with sequences bifurcating at the fourth–harmonic points of compressible Riemann sequences. Unfortunately, the parameter space is enormous and the stability analyses increase in complexity as higher harmonics are considered. In any case, it would presently be realistic to expect that the evolutionary path presented in this paper for viscous incompressible Jacobi ellipsoids does have an analog in the viscous compressible case as well. In fact, it would be surprising if the results of this paper were only applicable to strictly incompressible fluid–masses.

## 5   Discussion and Summary

In this paper, we have used the physical picture of phase transitions between equilibrium sequences of rotating, self–gravitating fluid–masses presented in Papers I and II in order to analyze the stability of incompressible Jacobi ellipsoids past the third–harmonic bifurcation point of the pear–shaped sequence. This physical picture has allowed us to identify some previous erroneous conclusions and to better understand the previous correct results.

The following previously obtained conclusions were found to be incorrect: (a) The Jacobi ellipsoids were believed to become both secularly and dynamically unstable beyond the third–harmonic Jacobi–pear bifurcation. (b) Tied to this, the Jacobi ellipsoids were believed to undergo cataclysmic evolution beyond the Jacobi–pear bifurcation. (c) It was also believed that, even in the case of evolution under conditions that maintain uniform rotation, secular and dynamical instability can appear together at the same point on an equilibrium sequence. (d) Finally, it was implicitly believed that performing linear stability analysis on the zero–vorticity Jacobi sequence has no differences compared to the corresponding analyses that are usually carried out on other equilibrium sequences of objects with vortical motions.

Our study has clarified the above issues and has led to the following physical interpretations of the results: (a) The Jacobi ellipsoids do not become secularly unstable beyond the Jacobi–pear bifurcation because the bifurcating pear–shaped sequence stands at higher energies and at higher rotation frequencies relative to the Jacobi sequence. This implies that the pear–shaped sequence is not accessible from the Jacobi sequence via quasistatic evolution driven by viscosity. Because of the absence of lower energy states, such secular stability further implies automatically dynamical stability and is in accordance with Kelvin's theorem of irrotational motion (e.g. Lamb 1932). (b) We conclude then that an evolving viscous Jacobi ellipsoid remains on its sequence beyond the third–harmonic bifurcation point. (c) In the case of phase transitions between sequences of uniformly rotating objects, secular and dynamical instability cannot occur at the same point or lead to the same sequence. This point has been discussed in §1.1 above. In the same context, the "dynamical instability" found by Cartan (1924) at the Jacobi–pear bifurcation cannot lead to the uniformly rotating pear–shaped objects. We have concluded that, in the complete absence of viscosity, a perfect–fluid Jacobi ellipsoid may be subject to a transition toward a vortical, differentially rotating state. However, a perturbed viscous Jacobi ellipsoid (whose equilibrium structure is not affected by viscosity) is not subject to Cartan's instability because vortical modes are destroyed by viscosity (cf. NGG) which also continually enforces a return back to the ellipsoidal form.
(d) This conclusion implies that linear stability analysis on the Jacobi sequence differs in one important detail from analogous analyses performed on other sequences of fluid–masses with vorticity: the normal modes of inviscid Jacobi ellipsoids must be constrained to be nonvortical. This leads to a different eigenvalue problem (cf. GGN) on the inviscid Jacobi sequence. The new eigenvalue problem is essentially two–dimensional and is not relevant in the physically interesting case of evolution driven by viscosity toward or along the Jacobi sequence (see §2.2).

We have shown in §3 that the above conclusions are confirmed by the numerical computations of Eriguchi, Hachisu, & Sugimoto (1982) who have obtained a pear–shaped sequence in the vicinity of the Jacobi–pear bifurcation. The numerical results are in agreement with the analytical results obtained by Jeans (1929) for slightly deformed pear–shaped objects near the same bifurcation point. Jeans has effectively found that pear–shaped objects exist at higher energies and at higher rotation frequencies compared to the nearby Jacobi ellipsoids. The numerical results of Eriguchi, Hachisu, & Sugimoto (1982), that are illustrated in Figures 1 and 2, show undoubtedly the same behavior.

The importance of the numerical results lies in the fact that they also indicate clearly the absence of a nonlinear $\lambda$–transition (Paper II) in the vicinity of the pear–shaped sequence. Added to the absence of secular and conventional dynamical instability, this result makes the pear–shaped sequence absolutely inaccessible from the Jacobi sequence. The complete absence of any kind of third–harmonic instability on the Jacobi sequence allows for fission to occur further up on the sequence. Specifically, an evolving viscous Jacobi ellipsoid will remain on its sequence even beyond the fourth–harmonic bifurcation point of the dumbbell–binary sequence (Paper II) and it will be finally subject to the secular $\lambda$–transition of type 2 that is associated with this bifurcating sequence. The type 2 $\lambda$–transition breaks the topology without breaking the symmetry of the ellipsoid and produces an equal–mass binary system. The breaking of topology allows a Jacobi ellipsoid to abandon its sequence and to evolve toward the binary sequence. This mechanism is the only way out of the sequence since viscous Jacobi ellipsoids are *not* susceptible to conventional secular and dynamical instabilities because they have zero vorticity (see §2.2). The "modified classical fission hypothesis" has been discussed in more detail in §4 along with other, modern approaches to the fission problem (cf. the review of Tassoul 1978).

We should point out that all the above results can be strengthened by some straightforward numerical computations. Toward this end, the pear–shaped sequence must be computed at better numerical resolution/accuracy. Also, the absolute stability of viscous Jacobi ellipsoids prior to the type 2 $\lambda$–point ($T/|W| = 0.20236$) and their breakup beyond this point can be checked by hydrodynamical simulations of incompressible fluids or even by simulations of compressible Jacobi ellipsoids of low degrees of compressibility. We should also mention for completeness that Bardeen et al. (1977) and Friedman & Schutz (1978a, b) did point out that the perturbations used in the formulation of the tensor–virial analysis (EFE) are problematic because they can potentially violate Kelvin's theorem (see §2.1) but did not extend their investigation to instabilities in viscous Jacobi ellipsoids.

So far in this work, we have not yet discussed the third–harmonic and fourth–harmonic points of bifurcation and instability along the Maclaurin sequence. This is because we considered the corresponding points on the Jacobi sequence more important due to their relevance to the fission problem. The higher–harmonic points of the Maclaurin sequence are, however, interesting and important in their own right, especially since most of them appear to be related to nonlinear $\lambda$–transitions and to the nonlinear resonances mentioned in §4.1. As we shall see in Paper IV, these nonlinear phase transitions hold some new and unexpected surprises that cannot be captured by linear stability analyses alone.

## Acknowledgments

We thank D. Brydges, C. McKee, and R. Narayan for stimulating discussions and I. Hachisu and N. Lebovitz for useful correspondence. IS is grateful to the Gauss Foundation for support and to K. Fricke, Director of Universitäts-Sternwarte Göttingen, for hospitality during a stay in which much of this work has been accomplished. IS thanks also the Center for Computational Studies of the University of Kentucky for continuing support. This work was supported in part by NASA grants NAGW–1510, NAGW–2447, NAGW–2376, and NAGW–3839, by NSF grant AST–9008166, and by grants from the San Diego Supercomputer Center and the National Center for Supercomputing Applications.